\documentclass[twocolumn,showpacs,amsmath,amssymb]{revtex4}

\usepackage{graphics}
\usepackage{dcolumn}
\usepackage{bm}
\usepackage{psfig}  
\usepackage{epsfig}

\begin{document}

\title{Coherent QCD multiple scattering in proton-nucleus collisions}

\author{Jianwei Qiu}
 \email{jwq@iastate.edu}

\affiliation{Department of Physics and Astronomy, 
Iowa State University, Ames, IA 50011, USA } 

\author{Ivan Vitev}
 \email{ivitev@lanl.gov}

\affiliation{Los Alamos National Laboratory, Theory Division and
Physics Division,  Los Alamos, NM 87545, USA }

\begin{abstract}
We argue that high energy proton-nucleus (p+A) collisions provide 
an excellent laboratory for studying nuclear size enhanced parton
multiple scattering where power corrections to the leading twist
perturbative QCD factorization approach can be systematically
computed. We identify and resum these corrections and calculate the 
centrality and rapidity dependent nuclear suppression of single 
and double inclusive hadron production at moderate transverse momenta.
We demonstrate that both spectra and dihadron correlations in p+A 
reactions are sensitive measures of such dynamical nuclear attenuation
effects.
\end{abstract}
                                               

\maketitle


Copious experimental data~\cite{QM2004} from central Au+Au 
reactions at the Relativistic Heavy Ion Collider (RHIC)
has generated tremendous excitement by pointing to the
possible creation of a deconfined state of QCD with energy 
density as high as 100 times normal nuclear matter 
density~\cite{Gyulassy:2004vg}. In order to diagnose its 
properties~\cite{Vitev:2002pf}, we first need to quantitatively 
understand the multiple scattering between a partonic probe 
and the partons of the medium in simpler strongly interacting 
systems, for example p+A. The importance of such theoretical 
investigations has been recently stressed by the measured intriguing 
deviation~\cite{Arsene:2004ux} of the moderate-$p_T$ single 
inclusive hadron spectra 
from the independent multiple nucleon-nucleon  
scattering limit~\cite{Vitev:2003xu} in d+Au reactions at RHIC.

In this Letter we present a systematic calculation
of the {\it coherent} multiple parton scattering with several 
nucleons in p+A collisions, which has so far been considered 
a challenge  for the perturbative QCD factorization 
approach~\cite{Collins:gx}. 
Coherent multiple scattering and the well-studied elastic multiple
scattering co-exist in nuclear collisions and their relative role 
depends on the probes (or observables) and the underlying dynamics.
Elastic scattering has played an important role in understanding 
the $k_T$-broadening and the Cronin effect~\cite{Vitev:2003xu}. 
It dominates when the probe is so localized that the quantum 
correlation between the different scattering centers can be neglected.
In terms of the factorization approach, contributions from 
coherent multiple scattering to a physical cross section are power
suppressed in comparison to the leading hard partonic 
processes~\cite{Qiu:2001hj}. However, as we demonstrate below,  
when enhanced by the large nuclear size these may become important 
at small and moderate transverse momenta at RHIC.


\begin{figure}[b!]
\begin{center}
\includegraphics[width=1.6in]{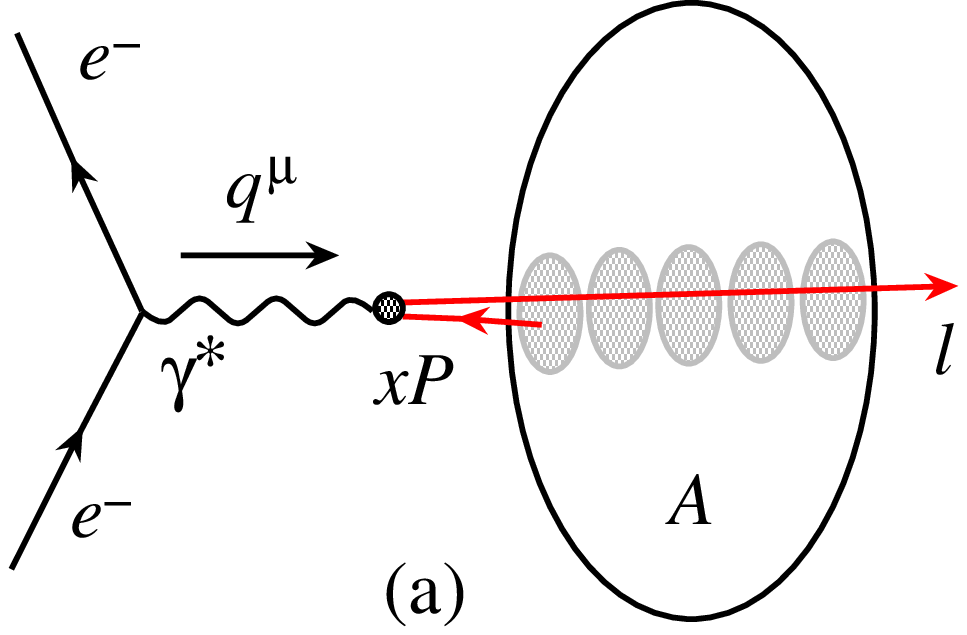}
\includegraphics[width=1.7in]{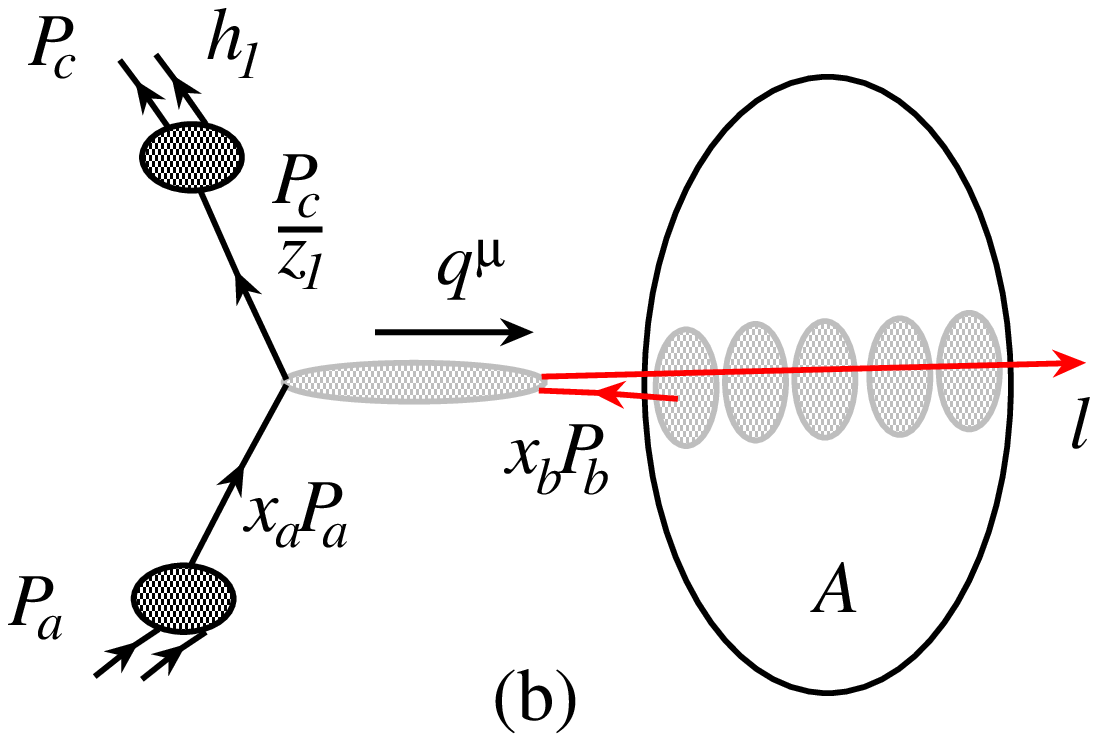}
\includegraphics[width=1.9in]{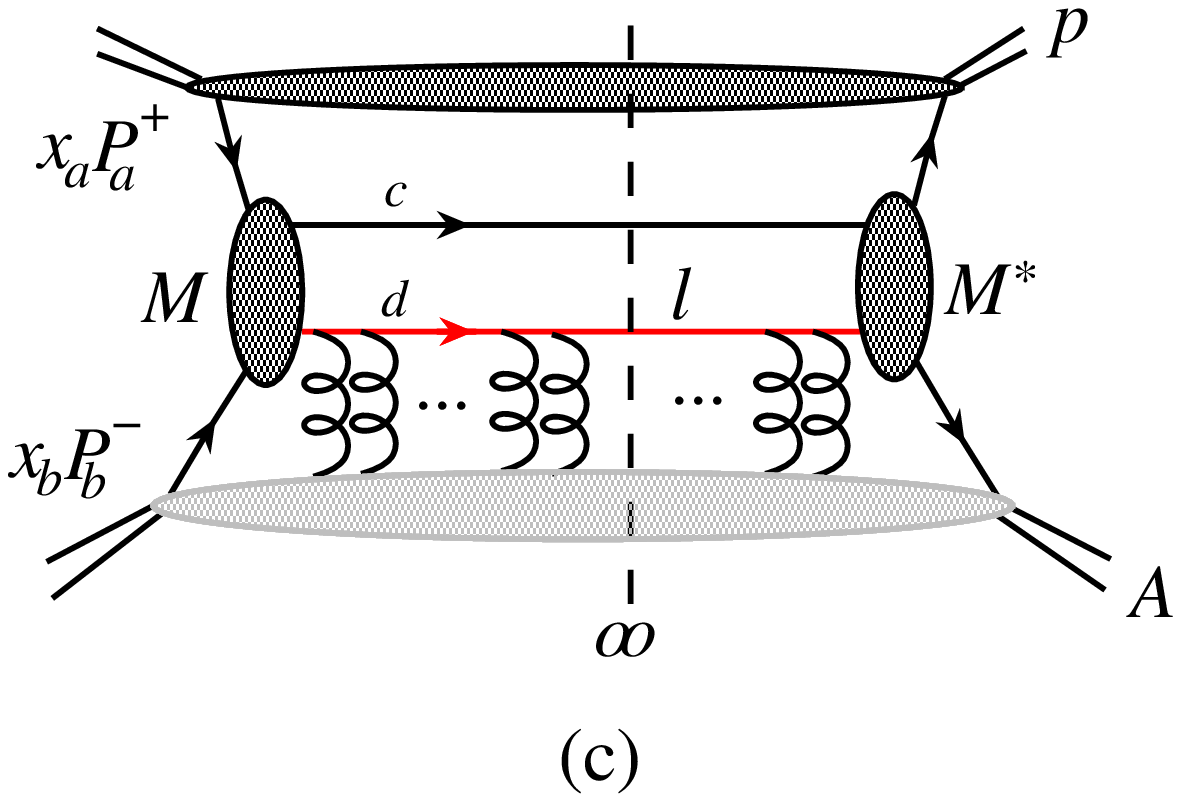}
\vspace*{-.1in}
\caption{Coherent multiple scattering of the struck parton 
in deeply inelastic scattering~(a) and in proton-nucleus 
collisions~(b). A set of resummed two gluon exchange 
diagrams~(c) in p+A reactions.}
\label{fig1}
\end{center} 
\vspace*{-4mm}
\end{figure}


Hard scattering in nuclear collisions requires one large momentum
transfer $Q \sim xP \gg \Lambda_{QCD}$ with parton momentum 
fraction $x$ and beam momentum $P$. A simple example, shown in 
Fig.~\ref{fig1}(a), is the lepton-nucleus deeply inelastic scattering (DIS).  
The effective longitudinal interaction length probed by the virtual photon
of momentum $q^\mu$ is characterized by $1/xP$. If the momentum fraction 
of an active initial-state parton $x \ll x_c=1/2m_N r_0\sim 0.1$  
with nucleon mass $m_N$ and radius  $r_0$, it could cover several  
Lorentz contracted nucleons  of longitudinal size $\sim 2r_0 (m_N/P)$
in a large nucleus~\cite{Qiu:2002mh}. In the photon-nucleus frame, 
Fig.~\ref{fig1}(a), the scattered parton of momentum $\ell$ 
will interact {\it coherently} with the partons from different nucleons 
at the same impact parameter~\cite{Qiu:2003vd}. The on-shell condition 
for $\ell$  fixes the initial-state parton's momentum fraction to 
the Bjorken variable $x = x_B=Q^2/(2P \cdot q)$ but additional 
final state scattering forces the rescaling 
$x = x_B(1+\xi^2 (A^{1/3}-1) /Q^2 )$~\cite{Qiu:2003vd}.
Here, $A$ is the  nuclear atomic weight and the scale of 
power corrections
\begin{equation}
\xi^2\approx \frac{3\pi\alpha_s(Q)}{8r_0^2}
\langle p|\hat{F}^2|p\rangle\;, 
\label{scale}
\end{equation} 
with the matrix element  $\langle p|\hat{F}^2|p\rangle = 
\frac{1}{2} \lim_{x\rightarrow 0}xG(x,Q^2)$.    
In Ref.~\cite{Qiu:2003vd} we demonstrated that the hard
parton interactions for any number of coherent multiple scattering 
in DIS are infrared safe, which explicitly verifies the factorization 
theorem. For  $\xi^2=0.09-0.12$~GeV$^2$ the calculated reduction 
in the DIS structure functions, known as nuclear shadowing, is 
consistent with the $x_B$-, $Q^2$- and $A$-dependence of the data 
and the small $Q^2$ modification of the QCD sum rules~\cite{Qiu:2003vd}. 
Coherence should play a similar role in virtual quark and gluon 
exchanges in hadronic collisions, as shown in Fig.~1(b).

Before we resum the power corrections in p+A reactions and derive 
the modification to the single and  double inclusive hadron 
production cross sections we need to address the 
validity of factorization in a hadronic environment. 
The colored soft gluon interactions between the incoming and/or 
outgoing hadrons can potentially ruin the 
factorization~\cite{Doria:ak,Basu:1984ba}. 
Nevertheless, factorization in 
hadronic collisions was proved to be valid at the leading 
power~\cite{Collins:gx} as well as the leading power corrections 
${\cal O}(1/Q^2)$~\cite{Qiu:xx}. 
It has been argued~\cite{Qiu:2003cg} that factorization can 
be extended  to the calculation of $(A^{1/3}/Q^2)^N$-type 
nuclear size enhanced power corrections in p+A
collisions at all powers of $N$.

In DIS, as shown in Fig.~\ref{fig1}(a), only diagrams with multiple
final-state interactions to the scattered parton lead to medium
size enhanced power corrections~\cite{Qiu:2003vd}.  On the other hand, 
in p+A collisions, shown in Fig.~\ref{fig1}(b), all
diagrams with either final-state and/or initial-state multiple
interactions could in principle contribute. 
However, the {\it coherent} multiple scattering is a result of 
an extended probe size, which is determined by the momentum exchange
of the hard collisions.  As shown in Fig.~\ref{fig1}(b), once we fix
the momentum fractions $x_a$ and $z_1$, the effective interaction region
is determined by the momentum exchange
$q^{\mu}=(x_aP_a-P_c/z_1)^{\mu}$.  In the head-on frame of $q-P_b$,
the scattered parton of momentum $\ell$ interacts coherently with
partons from different nucleons at the same impact parameter.
Interactions that have taken place between the partons from the 
nucleus and the incoming parton of momentum $x_aP_a$ and/or the 
outgoing parton of momentum $P_c/z_1$ at a different impact parameter are
much less coherent and actually dominated by the independent elastic
scattering~\cite{Gyulassy:2002yv}.


We pursue the analogy with the DIS results of Ref.~\cite{Qiu:2003vd} 
and express the lowest order single and double inclusive hadron 
production cross sections as follows:
\begin{widetext}
\begin{eqnarray} 
\label{single}
\frac{ d\sigma^{h_1 }_{NN} }{ dy_1  d^2p_{T_1} }  
& = &  \sum_{abcd}
\int \frac{dz_1}{z_1^2} D_{h_1/c}(z_1) 
\int d x_a   \frac{\phi_{a/N}(x_a)}{x_a}  
\left[\frac{1}{x_a S +  {U}/{z_1} }\right]  
\frac{ \alpha_s^2}{S}  
\int dx_b\,  \delta(x_b-\bar{x}_b) \, F_{ab\rightarrow cd}(x_b)  \, ,  
\\
\frac{ d\sigma^{h_1 h_2}_{NN} }{ dy_1  dy_2 d^2p_{T_1}  d^2p_{T_2} } 
&=& 
\frac{\delta (\Delta \varphi - \pi)}{p_{T_1} p_{T_2} } 
\sum_{abcd} 
\int \frac{dz_1}{z_1} \, D_{h_1/c}(z_1) \,
D_{h_2/d} (z_2)\,  \frac{\phi_{a/N}(\bar{x}_a)}{\bar{x}_a} \,
\frac{\alpha_s^2}{{S}^2 }
\int dx_b\,  \delta(x_b-\bar{x}_b) \,  F_{ab\rightarrow cd}(x_b) \, ,
\label{double}
\end{eqnarray}
\end{widetext} 
where $\sum_{abcd}$ runs over all parton flavors, $\phi_{a/N}(x_a)$,
$\phi_{b/N}(x_b)$ are the parton distribution 
functions (PDFs)~\cite{Gluck:1998xa} and $D_{h_1/c}(z_1)$, $D_{h_2/d}(z_2)$ 
are the fragmentation functions (FFs)~\cite{Binnewies:1994ju}. The dependence 
on the small momentum fraction $x_b$  is isolated in the function
\begin{equation}
F_{ab\rightarrow cd}(x_b) \equiv
\frac{\phi_{b/N}(x_b)}{x_b}\,  
|\overline {M}_{ab\rightarrow cd}|^2  \;, 
\label{xb-fun}
\end{equation}
where the $2 \rightarrow 2$ on shell squared partonic matrix elements 
$|\overline {M}_{ab\rightarrow cd}|^2$ are given in~\cite{Owens:1986mp}. 
In Eq.~(\ref{single})  the invariants $T= - p_{T_1}\sqrt{S}\, e^{-y_1}$, 
$U= - p_{T_1}\sqrt{S}\, e^{y_1}$  and the momentum fraction from the 
nucleus 
$\bar{x}_b = - T/(z_1 S + U/x_a)$.
In Eq.~(\ref{double}) $\Delta\varphi=\varphi_2-\varphi_1$, 
$z_2 = z_1\, p_{T2} / p_{T1}$, and the solution for the momentum 
fractions carried by the partons from the nucleon and the nucleus 
are $\bar{x}_a =  p_{T_1} ( e^{y_1} + e^{y_2}) / z_1 \sqrt{S}$ and  
$\bar{x}_b =  p_{T_1} ( e^{-y_1} + e^{-y_2}) / z_1 \sqrt{S} $,
respectively.  In both Eqs.~(\ref{single}) and (\ref{double})
the possible $K$-factor for the lowest order formula and all
renormalization and factorization scale dependences are 
suppressed. Ratios of cross sections in p+A and p+p reactions
are insensitive to their choice.


As in~\cite{Qiu:2003vd}, the leading $A^{1/3}$-enhanced 
contribution to the cross sections  comes from the coherent
scatterings with $N$ nucleons shown in Figs.~\ref{fig1}(b) 
and \ref{fig1}(c). The two gluon exchange with an {\em individual}
nucleon gives rise to the characteristic scale of high twist 
corrections $\xi^2=0.09 - 0.12$~GeV$^2$, defined in Eq.~(\ref{scale}).  
For fixed $N$, the final state interactions of parton $d$, illustrated in 
Fig~1(c), replace  the integral over the small $x_b$ dependent function  
$F_{ab\rightarrow cd}(x_b)$ in Eqs.~(\ref{single}) and (\ref{double}) by
\begin{eqnarray}
&&\int dx_b\; \frac{(-1)^N}{N!} 
     \left[ C_d\, \frac{\bar{x}_b\, \xi^2}{-t} (A^{1/3}-1) \right]^N
\nonumber \\[1ex]
& & {\hskip 1.in}
   \times \frac{d^{N} \delta(x_b - \bar{x}_b) }{dx_b^N} \; 
F_{ab\rightarrow cd}(x_b) \;.  \quad
\label{N-fix}
\end{eqnarray}
Here, the hard scale $t=q^2=(x_aP_a-P_c/z_1)^2$ and $C_d$ is a color 
factor depending on the flavor of parton $d$. $C_{q(\bar{q})}=1$ and
$C_g=C_A/C_F=9/4$ for quark (antiquark) and gluon, respectively.
By resumming all  contributions in Fig.~1(c), $N=0, \cdots, \infty$  
in Eq.~(\ref{N-fix}), 
we derive the nuclear modified cross sections for single and double 
inclusive hadron production in p+A collisions. These have the 
same form as Eqs.~(\ref{single}) and (\ref{double}) with
the following replacement:
\begin{eqnarray}
 F_{ab\rightarrow cd}(x_b) & \Rightarrow &
\exp \left[  C_d \frac{\xi^2}{-t} (A^{1/3}-1)\, 
x_b\frac{d}{dx_b} \right] \, F_{ab\rightarrow cd}(x_b) 
\nonumber \label{N-infty} \\
 \!\! = &&  \!\!\!\!\!
  F_{ab\rightarrow cd}\left(x_b \left[ 1+ C_d
\frac{\xi^2}{-t} (A^{1/3}-1) \right]\right)  \;.
\label{resum}
\end{eqnarray}
Similarly to the DIS case, resumming the coherent scattering
with multiple nucleons is equivalent to a shift of the momentum
fraction of the active parton from the nucleus and leads to a net
suppression of the cross sections. The $t$-dependence of this shift
in Eq.~(\ref{resum}) indicates that the attenuation increases 
in the forward rapidity region. 
%
   

\begin{figure}[t!]
\begin{center} 
\psfig{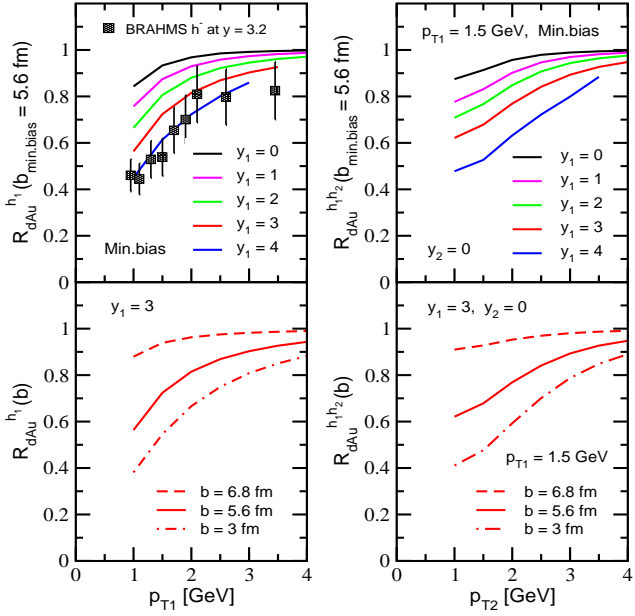}
\caption{Top: suppression of the single and  double inclusive 
hadron production rates in d+Au reactions versus $p_T$ for 
rapidities $y_1 = 0,1,2,3$ and $4$. $\xi^2 = 0.12$~GeV$^2$.  
Data on forward rapidity 
$h^-$ attenuation is from BRAHMS~\cite{Arsene:2004ux}. 
Bottom: impact parameter dependence of the calculated 
nuclear modification for central $b=3$~fm,
minimum bias $b_{\rm min.bias}=5.6$~fm and peripheral
$b=6.9$~fm collisions. The trigger hadron $p_{T1}=1.5$~GeV,   
$y_1 = 3$ and the associated hadron  $y_2=0$. }  
\label{fig2-mod-12}
\end{center} 
\end{figure}


We consider d+Au reactions at RHIC at $\sqrt{S}=200$~GeV with the deuteron 
and the nucleus moving with rapidities $y_{\max}=\pm 5.4$,  respectively.  
Small $p_{T_1} = 1$~GeV and forward $y_1=4$ hadron production 
corresponds to nuclear $x_b \geq 1.3 \times 10^{-4}$ where coherence 
and our calculated power corrections will certainly become important.   
Dynamical nuclear effects in multi-particle production can be 
studied through the ratio
\begin{equation}
\!  R^{(n)}_{AB}  =  \frac{d\sigma^{h_1 \cdots h_n}_{AB} / 
dy_1 \cdots dy_n d^2p_{T_1} \cdots d^2p_{T_n}} 
{\langle N^{\rm coll}_{AB} \rangle\, d\sigma^{h_1 \cdots h_n}_{NN} / 
dy_1 \cdots dy_n d^2p_{T_1} \cdots d^2p_{T_n}} \; .
\label{multi}
\end{equation}
Centrality dependence is implicit in Eq.~(\ref{multi}) and the 
modified cross section per average collision 
$d\sigma_{dAu}^{h_1 \cdots h_n}/\langle N^{\rm coll}_{dAu} \rangle $ 
can be calculated from   Eqs.~(\ref{single}) and (\ref{double}) 
via the substitution Eq.~(\ref{N-infty}). The necessary scaling 
with impact parameter, $b$, of the enhanced power corrections 
can be obtained through the nuclear thickness function,    
$A^{1/3} \rightarrow A^{1/3}T_A(b)/ T_A(b_{\rm min.bias})$, where
$T_A(b) = \int_{-\infty}^{\infty}\rho_A(z,b)dz$. 

The top panels of Fig.~\ref{fig2-mod-12} show the rapidity and
transverse momentum dependence of $R^{h_1}_{dAu}(b)$ 
and $R^{h_1h_2}_{dAu}(b)$ for minimum bias collisions. 
The amplification of the suppression effect at forward $y_1$ comes 
from the build-up of coherence
and the decrease of the Mandelstam variable 
$(-t)$. At high $ p_{T1},p_{T2} $ the attenuation is found to  
disappears in accord with the QCD factorization 
theorems~\cite{Collins:gx,Qiu:xx}. BRAHMS data~\cite{Arsene:2004ux} 
on $h^-$ production at forward rapidity is shown for comparison.  
Our result should be contrasted with a recent numerical study of 
non-linear parton evolution that found a factor of 2 suppression of 
the hadron production in p+A collisions at $p_{T} = 10^2 - 10^4$~GeV 
but negligible rapidity dependence of $R^{h_1}_{pA}(b)$ 
at small $p_{T} \leq 2$~GeV~\cite{Albacete:2003iq}.  
The bottom panels of Fig.~\ref{fig2-mod-12} show the growth of 
the  nuclear attenuation effect with centrality.

In the double collinear approximation, Eq.~(\ref{double}), the 
leading hadrons are produced strictly back-to-back. The acoplanarity, 
$\Delta \varphi  \neq \pi$~\cite{Angelis:1980bs}, arises from soft gluon 
resummation~\cite{Dokshitzer:1978hw}, next-to-leading order corrections
and in the presence of nuclear matter - 
transverse momentum diffusion~\cite{Gyulassy:2002yv}.  
We consider dihadron  correlations associated with  hard 
scattering partonic $2 \rightarrow 2$ 
processes that have the bulk many-body collision  background subtracted  
and can be approximated by near-side and away-side Gaussians:    
\begin{eqnarray}
C_2(\Delta \varphi) &=& \frac{1}{{\rm Norm}} 
\frac{dN^{h_1h_2}_{\rm dijet}}{d\Delta \varphi}
 \nonumber \\  
& \approx & \frac{A_{\rm Near}}{\sqrt{2\pi} \sigma_{\rm Near} }  
e^{-\frac{\Delta \varphi^2}{2\sigma^2_{\rm Near}} } + 
\frac{A_{\rm Far}}{ \sqrt{2\pi} \sigma_{\rm Far} } 
e^{-\frac{(\Delta \varphi -\pi)^2}{2\sigma^2_{\rm Far}} }. \qquad 
\label{cor-fun}
\end{eqnarray}
In Eq.~(\ref{cor-fun}) the near-side width $\sigma_{\rm Near}$ of  
$C_2(\Delta \varphi)$ is determined by jet fragmentation~\cite{Rak:2004gk}. 
If the strength of the away-side correlation function in elementary 
N+N collisions is normalized  to unity, dynamical quark and gluon 
shadowing  in p+A reactions will be manifest in the  attenuation of  
the {\em area}  $A_{\rm Far} = R^{h_1h_2}_{pA}(b)$. The away-side width 
$\sigma_{\rm Far}$ is related~\cite{Gyulassy:2002yv,Rak:2004gk} to 
the accumulated dijet  transverse momentum squared  
in a plane perpendicular to the collision axis:
\begin{equation}
\langle k_T^2 \rangle_{\rm dijet} = 2 \, \langle k_T^2 \rangle_{\rm vac} + 
\sum_{i=a,c,d}  \frac{ \mu^2 L_{\rm eff}}{\lambda}
(1 + \tanh^2 y_i) \;.
\label{projrct}
\end{equation}
In Eq.~(\ref{projrct})  $y_i$ are the interacting  parton rapidities 
and $\mu^2 L_{\rm eff} /\lambda = 0.72$~GeV$^2$  characterizes  
the elastic scattering power of cold nuclear matter in minimum bias 
d+Au collisions~\cite{Vitev:2003xu,Gyulassy:2002yv} and varies with 
centrality  $\sim T_A(b)/T_A(b_{\rm min.bias})$. The vacuum broadening 
$\langle k_T^2 \rangle_{\rm vac}$ is taken  from the 
data~\cite{Rak:2004gk,Adams:2003im}

\begin{figure}[t!]
\begin{center} 
\psfig{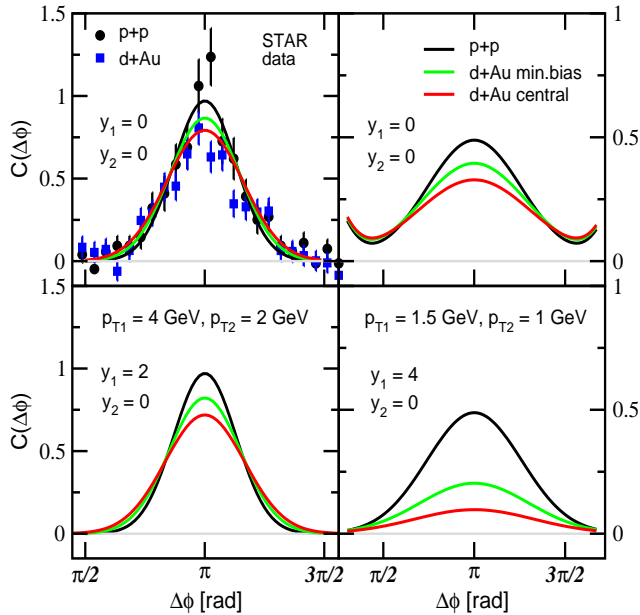}
\caption{Left: centrality dependence of 
$C_2(\Delta \varphi)$ at moderate  $p_{T_1}=4$~GeV, 
$p_{T_2}=2$~GeV and rapidities $y_1=0,2$ and $y_2=0$.  
Central d+Au and p+p data  from STAR~\cite{Adams:2003im}.  
Right: $C_2({\Delta \varphi})$ at small transverse momenta 
$p_{T_1}=1.5$~GeV, $p_{T_2}=1$~GeV and rapidities $y_1=0,4$ and $y_2=0$.}
\label{fig2-corr-fun}
\end{center} 
\end{figure}

In triggering on a large transverse momentum hadron,
the nuclear modification from enhanced power corrections and 
transverse momentum diffusion is reflected in the away-side
correlation function. For large vacuum broadening,  
$\langle k_T^2 \rangle_{\rm vac} = 2.5 - 3.5$~GeV$^2$, 
elastic scattering may lead to only a moderate additional growth
of $\sigma_{\rm Far}$. The left panels of Fig.~\ref{fig2-corr-fun} 
show that for $p_{T_1}=4$~GeV, $p_{T_2}=2$~GeV 
the dominant effect in $C_2(\Delta \varphi)$ is a small increase 
of the broadening with centrality, compatible with the 
PHENIX~\cite{Rak:2004gk}  and STAR~\cite{Adams:2003im} 
measurements. Even at forward rapidity, such as $y_1 = 2$, the 
effect of power corrections in this transverse momentum  range is  
not very significant. At small $p_{T_1}=1.5$~GeV,  $p_{T_2}=1$~GeV, 
shown in the right hand side of  Fig.~\ref{fig2-corr-fun}, the apparent 
width of the away-side  $C_2(\Delta \varphi)$ is larger. 
In going from  midrapidity,  $y_1 = y_2 = 0$, to forward  rapidity, 
$y_1 = 4, y_2 = 0$,  we find a significant reduction  by a factor 
of 3 - 4  in the strength
of dihadron correlations from the nuclear enhanced power corrections. 
We emphasize again that this result depends 
sensitively on centrality and  the choice of transverse momenta 
and disappears at high $p_T$.


The dynamical cross section attenuation calculated here does not
contradict, instead, complements the effects from a possible
modification of the nuclear parton distribution functions (nPDFs),
known as leading twist shadowing~\cite{Frankfurt:2003zd}. In our
formalism, one can include both effects by replacing the PDFs in
Eqs.~(\ref{single}) and (\ref{double}) with the corresponding nPDFs
and applying Eq.~(\ref{N-infty}).  The weaker scale dependence of 
nPDFs will thus slow down the disappearance of the 
nuclear suppression at high $p_T$ in Fig.~\ref{fig2-mod-12}.  
A combined global QCD analysis would, however, be required and 
is beyond the scope of this Letter.


In conclusion, we presented a systematic approach to the calculation
of coherent QCD multiple scattering and resummed the
nuclear enhanced power correction to the single and double inclusive 
hadron production cross sections in p+A reactions. At low $p_T$ 
we find a sizable  suppression, which grows with rapidity and  
centrality. At high $p_T$ the nuclear modification disappears 
in accord with the QCD factorization theorems. We demonstrated that 
both particle  spectra and dihadron correlations are sensitive 
measures of such dynamical attenuation  of the parton interaction
rates. 

Our approach, with its intuitive and transparent results, 
can be  easily applied to study the nuclear 
modification of other physical observables in p+A 
reactions and its predictions can be readily tested against 
RHIC data. The systematic incorporation of coherent power 
corrections provides a tool to address the most interesting 
transition  region between ``hard'' and ``soft'' physics
in hadron-nucleus collisions.  It allows for the extension 
of perturbative QCD calculations  to relatively small transverse 
momenta and for bridging the gap between the parton model 
picture and the possible onset of gluon 
saturation~\cite{Gyulassy:2004vg,McLerran:1993ka}
at very large collider energies and very small values 
of nuclear $x$.

This work is supported in part by the US Department of Energy  
under Grant No. DE-FG02-87ER40371 and by the J. Robert Oppenheimer
fellowship of the Los Alamos National Laboratory.  


\end{document}